\documentclass[prd,tightenlines,nofootinbib,superscriptaddress]{revtex4}

\usepackage{amsfonts,amssymb,amsthm,bbm,amsmath}
\usepackage{hyperref}

\usepackage{color}

\usepackage{tikz}
\usetikzlibrary{calc}
\usetikzlibrary{decorations.pathmorphing}
\usepackage{subcaption}
\newcommand{\C}{{\mathbb C}}
\newcommand{\N}{{\mathbb N}}
\newcommand{\R}{{\mathbb R}}

\newcommand{\cG}{{\mathcal G}}

\newcommand{\cH}{{\mathcal H}}

\newcommand{\cN}{{\mathcal N}}

\newcommand{\cO}{{\mathcal O}}

\newcommand{\cC}{{\mathcal C}}

\newcommand{\SL}{\mathrm{SL}}

\newcommand{\be}{\begin{equation}}
\newcommand{\ee}{\end{equation}}
\newcommand{\beq}{\begin{eqnarray}}
\newcommand{\eeq}{\end{eqnarray}}
\newcommand{\bes}{\begin{eqnarray}}
\newcommand{\ees}{\end{eqnarray}}

\renewcommand{\sl}{{\mathfrak{sl}}}

\newcommand{\la}{\langle}
\newcommand{\ra}{\rangle}

\newcommand{\f}{\frac}

\def\nn{\nonumber}
\def\pp{\partial}

\def\rd{\mathrm{d}}

\def\ka{\kappa}
\def\vphi{\varphi}
\def\eps{\epsilon}

\def\om{\omega}

\def\hH{\hat{H}}
\def\hX{\hat{X}}
\def\hP{\hat{P}}
\def\hV{\hat{V}}
\def\hx{\hat{x}}
\def\hp{\hat{p}}
\def\hD{\hat{D}}
\def\hQ{\hat{Q}}
\def\hcO{\hat{\cO}}
\def\hK{\hat{K}}
\def\bpsi{\bar{\psi}}
\def\hvsigma{\hat{\varsigma}}
\def\hc{\hat{c}}
\def\hC{\hat{\cC}}

\begin{document}

\title{Quantizing the Quantum Uncertainty}
%\title{Conformal symmetry, Effective dynamics \& Quantization of the quantum uncertainty}

\author{{\bf Etera R. Livine}}\email{etera.livine@ens-lyon.fr}
\affiliation{Laboratoire de Physique, ENS Lyon, CNRS-UMR 5672, 46 all\'ee d'Italie, Lyon 69007, France}
%\affiliation{Perimeter Institute for Theoretical Physics,  Waterloo, Ontario N2L 2Y5, Canada}

\date{\today}

\begin{abstract}

The spread of the wave-function, or quantum uncertainty, is a key notion in quantum mechanics. At leading order, it is characterized by the quadratic moments of the position and momentum operators. These evolve and fluctuate independently from the position and momentum expectation values. They are extra degrees of quantum mechanics compared to classical mechanics, and encode the shape of wave-packets.
Following the logic that quantum mechanics must be lifted to quantum field theory, we discuss the quantization of the quantum uncertainty as an operator acting on wave-functions over field space and derive its discrete spectrum, inherited from the $\sl_{2}$ Lie algebra formed by the operators $\hx^{2}$, $\hp^{2}$ and $\widehat{xp}$.
We further show how this spectrum appears in the value of the coupling  of the effective conformal potential driving the evolution of extended Gaussian wave-packets according to Schr\"odinger equation, with the quantum uncertainty playing the same role as an effective intrinsic angular momentum.
% for the quadratic moments.
%
We conclude with an open question: is it possible to see experimental signatures of the quantization of the quantum uncertainty in non-relativistic physics, which would signal the departure from quantum mechanics to a QFT regime?

\end{abstract}

\maketitle
\tableofcontents
%%%%%%%%%%%%%%%%%%%%%%%%%%%%%%%%%%%%%%%%%%%%%%%%%%%%%%%%

%%%%%%%%
\section*{Introduction}
%%%%%%%%

Quantum mechanics is a huge leap in our understanding of physics compared to classical mechanics. Its rich phenomenology, especially reformulated in terms of quantum information and multipartite entanglement, is at the heart of nowadays' quantum technology revolution. We know however that it needs to be upgraded to quantum field theory. Although one might think at first about relativistic quantum field theories and gauge field theories, as used in high energy particle physics, non-relativistic quantum field theories are also crucial to condensed matter, nuclear physics, and even to understanding subtle points of quantum information (see e.g. \cite{Anastopoulos:2023khd}).

Following this logic, working in one spatial dimension for the sake of mathematical simplicity, one starts with 1d classical mechanics, with trajectories $x(t)$ describing the evolution of a point-like object in time. This gets lifted up to 1d quantum mechanics, with an evolving wave-function $\psi(t,x)$ giving the probability amplitude density of the system being at the position $x$ at the time $t$. The wave-function $\psi$ is in fact a (non-relativistic) field over space-time. In turn, this should be get 2nd-quantized and lifted to 1d Schr\"odinger quantum field theory, with a wave-function over the space of fields $\Psi[\psi]$. 

Let us nevertheless remember that quantization is not a physical process, but consists in mathematical methods carefully designed and crafted to guess  more fundamental theories with a broader regime of validity (for instance, valid at smaller length scales) from effective theories valid in a more restricted range of physical parameters. From this perspective, there is no actual real distinction between a first or a second quantization, except for a pedagogical effort to explain how to build quantum mechanics and quantum field theory step by step.

The opposite mathematical procedure -``dequantizing''- is a more physically-motivated process of extracting an effective mechanical description from the more fundamental theory, and in particular understanding how quantum systems can behave as classical objects in certain regimes. This is usually pictured as taking a $\hbar\rightarrow 0$ limit, but using only the Planck constant as marker blurs the line between quantum mechanics and quantum field theory. Indeed,  setting a $\hbar\rightarrow 0$,  would  one descend directly from quantum field theory to classical mechanics, or at which point would one go through an intermediate quantum mechanical regime?
%\footnotemark{}? 
%
%\footnotetext{
%This echoes a more general question: when is a physical phenomenon to be considered classical and not quantum? All coupling constants in chemistry or physics have a microscopic origin involving quantum physics, so should they be considered as quantum? In a similar fashion, in some case, the quantum uncertainty can be considered as having a classical dynamics.
%}

In this context, we would like to revisit the notion of quantum uncertainty and investigate its fate in quantum field theory.
Quantum uncertainty is the foremost quantum phenomenon, before entanglement or decoherence. It is not an anomaly for classical mechanics or a to-be-suppressed noise around the classical motion, quantum mechanics turns it into an actual degree of freedom, which stands on its own. It is a physical quantity, that can be measured (e.g. the energy spectrum of the harmonic oscillator) and experimentally modulated (e.g. wave-packet squeezing).
It is mathematically quantified at leading order by the Robertson-Schr\"odinger inequality. Considering the quadratic moments of the wave-function, defined as the expectation values if the quadratic polynomials in the position and momentum, the Heisenberg uncertainty principle can be written as:
\be
c\equiv\Big{[}\la x^{2}\ra-\la x\ra^{2}\Big{]}\Big{[}\la p^{2}\ra-\la p\ra^{2}\Big{]}-\Big{[}\la xp\ra-\la x\ra\la p\ra\Big{]}^{2}
\,,\qquad
c\ge\f{\hbar^{2}}4
\,.
\ee
The observable $c$, and more generally the quadratic moments, describe the spread and shape of the wave-function. 

The evolution of the linear and quadratic moments, $\la x\ra$, $\la p\ra$, $\la x^{2}\ra$, $\la p^{2}\ra$, $\la xp\ra$, and thus of the Robertson-Schr\"odinger uncertainty $c$, is dictated by Schr\"odinger equation. In fact, one can extract effective equations of motion for all linear, quadratic and higher order, moments of the wave-functions.
This supports the semi-classical description of a quantum particle as  a classical object dressed with extra quantum degrees of freedom \cite{ABCV,Prezhdo2000QuantizedHD,Baytas:2018ruy,Baytas:2018gbu,Bojowald:2022lbe}. These degrees of freedom are the higher moments, which carry the information on the shape of the wave-function and its fluctuations around its classical position and momentum. This perspective allows for an efficient description of quantum phenomenology in semi-classical terms \cite{Prezhdo2000QuantizedHD} (see also \cite{,Elze:1994tg,Blum:1995wi,Jalabert_2001,Dong:2022dfo} for applications to the effective evolution of  open quantum systems).

This point of view has been extensively developed in the framework of the dynamics of Gaussian wave-packets \cite{ABCV,Prezhdo2000QuantizedHD}. Indeed, reducing the wave-function data to a finite number of degrees of freedom encoded by its quadratic moments allows to write the  dynamics of Gaussian states as an effective classical Hamiltonian mechanics \cite{Baytas:2018ruy,Baytas:2018gbu,Bojowald:2022lbe,Jackiw:1979wf,Heslot:1985qam}, with an effective potential driving the diffusion of the wave-packets and whose coupling constant is the Robertson-Schr\"odinger uncertainty $c$. Relevant for both theory and experiment, this provides an efficient approximation of the quantum dynamics in the semi-classical regime\footnotemark{}.
%, as long as multi-body entanglement can be neglected.
%
\footnotetext{
One can go beyond the Gaussian approximation by including the higher moments of the wave-function. Schr\"odinger equation leads to a coupled evolution for all the moments, where higher moments can in general interact with lower moments and affect the classical trajectories. The evolution of the whole tower of higher moments is more tricky to describe than the restriction to quadratic moments. Although it is possible to integrate the whole evolution explicitly for a time-dependent harmonic potential (see e.g. \cite{Livine:2023vph}), it remains a challenge to write the evolution of higher moments down as an effective Hamiltonian evolution for an arbitrary potential \cite{Baytas:2018ruy,Baytas:2018gbu}.
}

Now having a classical mechanics formulation of Gaussian wave-packets naturally leads to the question: could one quantize this dynamics? On the one hand, asking if one should quantize the semi-classical degrees of freedom characterizing the Gaussian approximation of a wave-packet is clearly a lopsided question. On the other hand, putting aside the Gaussian wave-packets and considering the general dynamics of arbitrary wave-functions, the quadratic moments, such as the position spread $\la x^{2}\ra=\int \rd x\,x^{2}|\psi|^{2}$, are field observables on the same level as the wave-function Fourier modes $\psi_{p}=\int \rd x\, e^{-ipx}\psi$.
The second quantization, upgrading quantum mechanics to quantum field theory, should thus raise all the wave-function moments, $\la x\ra$, $\la p\ra$, $\la x^{2}\ra$, and so on, to  (Hermitian) operators acting on field wave-functionals $\Psi[\psi]$. The ensuing question is then: what should be the fate of the quantum uncertainty in quantum field theory? Or, in more technical terms, what would  be the spectrum of the quantum uncertainty operator $\hat{c}$? What would be the meaning of its eigenvectors in the first quantized setting of quantum mechanics? For instance, could one see any consequence on the effective Hamiltonian dynamics of wave-packets?

\medskip

The first section of this short paper is dedicated to defining the uncertainty and spreads of the wave-functions and reviewing their evolution and their effective Hamiltonian dynamics. In particular, we recall the $\sl_{2}$ Lie algebra formed by the quadratic operators  $\hx^{2}$, $\hp^{2}$ and $\widehat{xp}$, and show how it allows to integrate the equation of motion for the position spread.
The second section investigates the quantization of the quantum uncertainty in quantum field theory. Using the $\sl_{2}$ algebraic structure, we  show that its spectrum is discrete and given by the $\sl(2,\R)$ Casimir operator for unitary representations. Field wave-functions, or wave-functionals, $\Psi[\psi]$ can be organized in terms of their degree in $\psi$. This degree is the eigenvalue of the number of particle operators. One-particule states are smearings of the wave-function $\psi$, two particle states are 2-point correlations of $\psi$ and so on. We compute explicitly the action of the quantum uncertainty operator on 1-, 2- and 3- particle states, writing it in terms of an angular momentum operator and confirming the discrete spectrum.
Finally, in the third section, we identify  squeezed coherent states as the ``dequantized'' version of the quantum uncertainty eigenstates, or more precisely the shadow in quantum mechanics of the QFT spectrum, in the similar sense that Bohr-Sommerfeld orbits are the shadow in classical mechanics of the spectrum of the Hamiltonian operator.

%%%%%%%%
%\section{$\sl_{2}$ symmetry \& the quantum uncertainty algebra}
\section{The quantum uncertainty algebra}
%\section{The quantum uncertainty $\sl_{2}$ algebra}
%%%%%%%%

An essential simple algebraic structure in mechanics, just beyond the well-known canonical Poisson bracket between position and momentum leading to the canonical commutator in quantum mechanics,
\be
\{x,p\}=1
\quad\Rightarrow\quad
[\hx,\hp]=i\hbar
\,,
\ee
is the $\sl_{2}(\R)$ Lie algebra formed by the quadratic polynomials of the position and momentum,
\be
\label{eqn:sl2}
\left|\begin{array}{lcl}
\{x^{2},p^{2}\}&=&+4D
\,,\\
\{D,p^{2}\}&=&+2p^{2}
\,,\\
\{D,x^{2}\}&=&-2 x^{2}
\,,
\end{array}
\right.
\quad\textrm{with}\,\, D=xp
%\,.\ee\be
\qquad\Rightarrow\quad
\left|\begin{array}{lcl}
{[}\hx^{2},\hp^{2}{]}&=&+4i\hbar\hD
\,,\\
{[}\hD,\hp^{2}{]}&=&+2i\hbar\hp^{2}
\,,\\
{[}\hD,\hx^{2}{]}&=&-2i\hbar\hx^{2}
\,,
\end{array}\right.
\quad\textrm{with}\,\, \hD=(\hx\hp)_{sym}=\f12(\hx\hp+\hp\hx)
\,.
\ee
In classical mechanics, these quadratic polynomials are simply the generators of the $\SL(2,\R)$ linear canonical transformations in phase space,
\be
\delta \left(\begin{array}{c} x \\ p \end{array}\right)
=
\left\{
a\f{x^{2}}2+b\f{p^{2}}2+cxp
,\left(\begin{array}{c} x \\ p \end{array}\right)
\right\}
=
\left(\begin{array}{rr} -c & -b \\ a & c\end{array}\right)\left(\begin{array}{c} x \\ p \end{array}\right)
\,,\quad\textrm{with}\,\,
\left(\begin{array}{rr} -c & -b \\ a & c\end{array}\right)
\in\sl(2,\R)
\,.
\ee
In quantum mechanics, these quadratic operators control the squeezing modes of wave-packets and pulses (se e.g. \cite{Zelaya:2021vxv,Gazeau:2022pzz}), and, from a more theoretical perspective, generate conformal transformations of the wave-functions (see e.g. \cite{Niederer:1972zz}). Then, in quantum field theory, they lead to the Bogoliubov transformations, widely used in quantum phenomenology, for instance in the description of supraconductivity or in the derivation of the Unruh effect.
It further appears in the foundation of 2-time physics \cite{Bars:1998pc}, in the quantization of the conformal particle \cite{deAlfaro:1976vlx}, in the identification of conformal symmetries in gravitational systems \cite{BenAchour:2023dgj} and cosmological models \cite{BenAchour:2020njq}, and so on.

Here, we would like to underline the role of this algebra in the dynamics of the quadratic spread of the wave-function, and thus in driving the evolution of the quantum uncertainty.

%%%
\subsection{Evolution of Uncertainty}
\label{sec:evolution}
%%%

For the sake of simplicity, let us work with 1d quantum mechanics and study the evolution of a wave-function $\psi(x)$ (in the position polarization) whose dynamics is defined by a linear Schr\"odinger (without non-linear self-interaction terms),
\be
i\hbar \pp_{t}\psi
=
\hH \psi
=
-\f{\hbar^{2}}{2m}\pp_{x}^{2}\psi
+\hV\psi
\,,
\ee
where the potential operator is typically chosen
%(but not necessarily)
as solely depending on the position, $\hV \psi(x)=V(x)\psi(x)$ in terms of a classical potential $V(x)$.
Using the standard Schr\"odinger representation for the position and momentum operators, 
\be
\hx=x\times
\,,\qquad
\hp=-i\hbar\pp_{x}
\,,\qquad
[\hx,\hp]=i\hbar
\,,
%\qquad\hH=\f1{2m}\hp^{2}+V(\hx)\,,
\ee
the evolution of observables is given by commutator with Hamiltonian, indeed for an initial quantum state $\psi_{0}$:
\be
\la \psi|\hcO|\psi\ra=\la \psi_{0}|e^{\f{+i}\hbar\hH t}\,\hcO\,e^{\f{-i}\hbar\hH t}|\psi_{0}\ra
\,,\qquad
\rd_{t}\la \psi|\hcO|\psi\ra=
\f i\hbar\la \psi|\,[\hH,\hcO]\,|\psi\ra
\,.
\ee
Applying this to the position and momentum operators, the standard computation gives:
\be
\la x\ra=\la \psi|\hx|\psi\ra
\,,\qquad
\rd_{t}\la x\ra
=
\f{-i}\hbar\la \psi|[\hx,\hH]|\psi\ra
=
\f1m\la \psi|\hp|\psi\ra
=
\f1m\la p\ra\,,
\ee
\be
\rd_{t}\la p\ra
=
\f{-i}\hbar\la \psi|[\hp,\hH]|\psi\ra
=
\f{-i}\hbar\la \psi|[\hp,V(\hx)]|\psi\ra
=
-\la \widehat{\pp_{x}V(x)}\ra\,.
\ee
This almost reproduces the exact classical equations of motion.
The deviation from classical dynamics is encoded in the fact that the effective potential term above $-\la \widehat{\pp_{x}V(x)}\ra$ is not exactly the classical force $-\pp_{x}V(\la x\ra)$ (evaluated on the mean position) and depends on the state $\psi$. When the potential is quadratic in $x$ and thus the force linear, there is actually no difference between the dynamics of the expectation values $\la x\ra $ and $\la p \ra$ and the classical trajectories. Quantum corrections creep in as soon as higher moments are involved, with terms in $\la x^{2}\ra$ and higher powers of both $x$ and $p$. The quadratic moments, $\la x^{2}\ra$, $\la p^{2}\ra$ and $\la xp\ra$, define the leading order spread of the wave-function  uncertainty, and thus the quantum uncertainty,  and higher moments can be understood as ``quantum hair'' dressing the motion of the classical particle. Indeed, quadratic and higher moments are not determined by the linear expectation values. They are extra degrees of freedom of  quantum systems, they evolve on their own and couple to the linear expectation values $\la x\ra $ and $\la p \ra$ and to the evolution of higher moments.

Let us then compute the commutator of the quadratic operators with the Hamiltonian and get the evolution equations of the quadratic moments:
\be
%[\hx^{2},\hH]
%=
%\f{2i\hbar}{m}\hD
%\quad\Rightarrow\quad
\rd_{t}\la x^{2}\ra
=\f2{m}\la \hD \ra
\,, \quad
%\ee
%\be
%\f12[\hx\hp+\hp\hx,\hH]
%%=
%%\f1{4m}[\hx\hp+\hp\hx,\hp^{2}]
%=
%\f1{4m}\left(
%[\hx,\hp^{2}]\hp
%+
%\hp[\hx,\hp^{2}]
%\right)
%+
%\f12(\hx[\hp,V(\hx)]+[\hp,V(\hx)]\hx)
%=
%\f{i\hbar}{m}\hp^{2}
%-
%i\hbar x\pp_{x}V
%\,,
%\ee
%\be
%\Rightarrow\quad
\rd_{t}\la \hD\ra
=
\f1{m}\la p^{2} \ra-\la x\pp_{x}V\ra
\,,
%\ee
%\be
%[\hp^{2},\hH]
%=
%[\hp^{2},V(\hx)]
%=
%-i\hbar\left(
%\hp\pp_{x}V+\pp_{x}V\hp
%\right)
%\quad\Rightarrow
\quad
\rd_{t}\la p^{2}\ra
=
-2\la \widehat{p\pp_{x}V}\ra_{sym}
\,,
\ee
where we recall that $\hD=\f12(\hx\hp+\hp\hx)$ is the generator of dilatations. When the potential includes cubic or higher terms in the position,  these equations clearly couple the quadratic moments to higher momenta of the wave-function. On the other hand, for a quadratic potential, i.e. a harmonic potential, these evolution equations form a closed set of differential equations, which can be straightforwardly integrated.
For instance, in the free case for a vanishing potential, these equations reduce to:
%Second order evolution:
\be
\label{eqn:freeevolve}
\rd_{t}\la \hx^{2}\ra=\f 2m\la \hD \ra
\,,\qquad
\rd_{t}\la \hD\ra=\f 1m\la \hp^{2} \ra
\,,\qquad
\rd_{t}\la \hp^{2} \ra=0\,.
\ee
The  momentum spread $\la \hp^{2} \ra$ is conserved and the position spread irremediably evolves as a quadratic function of time,
\be
\la \hx^{2}\ra=\f{\la \hp^{2} \ra}{m^{2}}(t-t_{0})^{2}+X_{0}\,,
\ee
where $t_{0}$ and $X_{0}$ are two constants of integration. Since the momentum spread $\la \hp^{2} \ra$ is strictly positive, this means that the position spread $\la \hx^{2}\ra$ will always grow at late times, illustrating the ineluctable diffusive behavior of the free Schr\"odinger equation.

For a harmonic potential, $V=\f12m\om^{2}x^{2}$, with constant frequency $\om$, the momentum spread is not conserved anymore and the evolution equations read:
\be
\label{eqn:harmonicevolve}
\rd_{t}\la \hx^{2}\ra=\f 2m\la \hD \ra
\,,\qquad
\rd_{t}\la \hD\ra=\f 1m\la \hp^{2} \ra -m\om^{2}\la \hx^{2}\ra
\,,\qquad
\rd_{t}\la \hp^{2} \ra=-2m\om^{2}\la \hD\ra
\,,
\ee
or equivalently in terms of 2nd order differential equation,
\be
\rd_{t}^{2}\la \hx^{2}\ra+2\om^{2}\la \hx^{2}\ra=\f 2{m^{2}}\la \hp^{2}\ra
\,,\qquad
\rd_{t}^{2}\la \hp^{2}\ra+2\om^{2}\la \hp^{2}\ra=2m^{2}\om^{4}\la \hx^{2}\ra
\,.
\ee
Following for example the general resolution for all higher moments of the wave-function for a time-dependent harmonic potential presented in \cite{Livine:2023vph}, we identify the solutions as oscillators in twice the frequency (as expected):
\be
\la \hx^{2}\ra=X_{0}\cos2\om(t-t_{0})
\,,\quad
\la \hp^{2}\ra=-m^{2}\om^{2}X_{0}\cos2\om(t-t_{0})
\,,
\ee
illustrating that the harmonic potential bounds the evolution of the position spread and allows for localized wave-packet solutions.

%\medskip

Now, whether we work with the free Schr\"odinger equation, and its purely diffusive behavior, or with a harmonic potential, and its localized behavior, the Hamiltonian $\hH$ is, in both cases, an element of the $\sl_{2}(\R)$ Lie algebra generated by the operators $\hx^{2}$, $\hp^{2}$ and $\hD$, so the following combination of moments is always a constant of motion:
\be
C=\left[
\la \hp^{2}\ra\la \hx^{2}\ra-\la \hD\ra^{2}
\right]\,,\qquad
\rd_{t}C
=0\,.
\ee
In order to make the dynamics of the spread and uncertainty more transparent, it is better to subtract the first order means and look directly at the quadratic fluctuations in position and momentum,
%Let us look at the uncertainty in position and momentum:
%\beq
%&&\sigma_{x}^{2}
%=
%\la \psi| (\hx -\la \hx\ra)^{2} |\psi\ra
%=
%\la x^{2}\ra-\la x \ra^{2}
%\nn\\
%&&\sigma_{p}^{2}
%=
%\la \psi| (\hp -\la \hp\ra)^{2} |\psi\ra
%=
%\la p^{2}\ra-\la p \ra^{2}
%\nn\\
%&&\sigma_{D}
%=
%\la D\ra-\la x \ra\la p \ra
%\eeq
\be
\sigma_{x}^{2}
=
\la x^{2}\ra-\la x \ra^{2}
\,,\quad
\sigma_{p}^{2}
=
\la p^{2}\ra-\la p \ra^{2}
\,,\quad
\sigma_{D}
=
\la D\ra-\la x \ra\la p \ra
\,,
\ee
and consider the combination:
\beq
c
&\equiv&
\la (\hp-\la \hp\ra)^{2}\ra\la (\hx-\la \hx\ra)^{2}\ra
-
\left[\f12\la (\hp-\la \hp\ra)(\hx-\la \hx\ra)+(\hx-\la \hx\ra)(\hp-\la \hp\ra)\ra\right]^{2}
\nn\\
&=&
\sigma_{x}^{2}\sigma_{p}^{2}-\sigma_{D}^{2}\,.
\eeq
%Define proper uncertainty:
%\beq
%c
%&=&
%\la (\hp-\la \hp\ra)^{2}\ra\la (\hx-\la \hx\ra)^{2}\ra
%-
%\left[\f12\la (\hp-\la \hp\ra)(\hx-\la \hx\ra)+(\hx-\la \hx\ra)(\hp-\la \hp\ra)\ra\right]^{2}
%\\
%&=&
%\la (\hp-\la \hp\ra)^{2}\ra\la (\hx-\la \hx\ra)^{2}\ra
%-
%\left[\la \hD\ra-\la \hx\ra\la \hp\ra\right]^{2}\,.
%\eeq
As well-known, applying the Cauchy-Schwarz inequality to the operators $\hx$ and $\hp$ leads to the Schr\"odinger inequality, which is a slight refinement of Heisenberg uncertainty relation:
\be
c\ge \f{\hbar^{2}}4
\,.
\ee
For a vanishing or harmonic potential, it is straightforward to check that this uncertainty is a constant of motion:
\be
\rd_{t}c=0\,.
\ee
It turns out that it is the coupling constant of the effective potential, of purely quantum origin, which drives the evolution of the positon and momentum uncertainties.
%the evolution equations of the quadratic moment is such that the momentum uncertainty def position uncertainty

In order to make this explicit, we introduce a pair of variables $(\alpha,\beta)$ encoding the spread of the wave-function
\be
\alpha=\sigma_{x}
\,,\quad
\alpha\beta=\sigma_{D}
\,,
\ee
so that the definition of $c$ gives the momentum uncertainty:
\be
\la x^{2}\ra=\la x \ra^{2}+\alpha^{2}
\,,\quad
\la p^{2}\ra=\la p \ra^{2}+\beta^{2}+\f{c}{\alpha^{2}}
\,.
\ee
For the free Schr\"odinger dynamics, with a vanishing potential $V=0$, one easily translates the equations \eqref{eqn:freeevolve} for the evolution of the quadratic moments into effective equations of motion for the pair $(\alpha,\beta)$:
\be
\label{eqn:effmotion}
\left|
\begin{array}{l}
\rd_{t}\la \hx^{2}\ra=\f 2m\la \hD \ra
\,,\vspace*{1mm}\\
\rd_{t}\la \hD\ra=\f 1m\la \hp^{2} \ra
\,,\vspace*{1mm}\\
\rd_{t}\la \hp^{2} \ra=0\,,
\end{array}\right.
\quad\Longrightarrow\quad
\left|
\begin{array}{l}
\rd_{t}c=0
\,,\vspace*{1mm}\\
\rd_{t}\alpha=\f \beta m\,,\vspace*{1mm}\\
\rd_{t}\beta=\f c m\f1{\alpha^{3}}\,.
\end{array}\right.
\ee
We recognize a Hamiltonian evolution, where $\beta$ is the conjugate momentum to $\alpha$, with an effective Hamiltonian\footnotemark,
\be
\label{eqn:effH}
\{\alpha,\beta\}_{eff}=1
\,,\quad
H^{(2)}_{eff}=\f{\beta^{2}}{2m}+\f{c}{2m\alpha^{2}}\,,
\ee
with a potential in $\alpha^{-2}$ with coupling $c>0$. Such a potential is identified as a conformal potential \cite{deAlfaro:1976vlx} and is similar to the effect of  an intrinsic 3d angular momentum generated of purely quantum origin.
Let us keep in mind that the value of the coupling $c$ is a priori not a universal coupling but depends on the initial  wave-function.
\footnotetext{
If we include a harmonic potential $V=\f12m\om^{2}x^{2}$, then a straightforward calculation from \eqref{eqn:harmonicevolve} shows that the effective Hamiltonian $H^{(2)}_{eff}$ for the quadratic uncertainty also inherits an extra copycat harmonic potential term in $V^{(2)}_{eff}=\f12m\om^{2}\alpha^{2}$.
}

Although the identification of the effective Hamiltonian driving the evolution of the quadratic moments, might seem ad hoc, as derived above, it can be derived in a more systematic way, directly from the field theory action for the Schr\"odinger equation, as we show below in the next section.

%%%
\subsection{Field theory and effective quantum mechanics}
\label{sec:Gaussian}
%%%

The evolution in time of a 1d wave-function according to Schr\"odinger equation is well understood as a non-relativistic 1+1-d field theory, with a field action:
\be
\label{eqn:Schroaction}
S[\psi]=\int \rd t\,\rd x\,\left[
i\hbar\bpsi\pp_{t}\psi
-\f{\hbar^{2}}{2m}\pp_{x}\bpsi\pp_{x}\psi
-V(x)|\psi|^{2}
%-\sum_{n\ge1}\f{\ka_{n}}n|\psi|^{2(n+1)}
\right]
\,,
\ee
where one recognizes the kinetic term $i\bpsi\pp_{t}\psi$, the dispersive term $\pp_{x}\bpsi\pp_{x}\psi$ and the potential term in $V(x)|\psi|^{2}$.
Not only this opens the door to investigating  non-linear effects by the addition of self-interaction terms to the action, in $\sum_{n\ge1}\f{\ka_{n}}n|\psi|^{2(n+1)}$, and subsequently second-quantizing and studying the ensuing non-trivial renormalization flow (see e.g. \cite{GUTKIN19881} for the quantization of the non-linear Schr\"odinger equation with $|\psi|^{4}$ potential), but this also underlines the degrees of freedom of the theory.

Indeed, comparing the Schr\"odinger field theory action above with the action for classical mechanics,
\be
s[x]=\int \rd t \,\left[
\f m2(\rd_{t}x)^{2} -V(x)
\right]
\,,
\ee
it is clear that quantum mechanics upgrades the one classical degree of freedom $x$ evolving into a trajectory $x(t)$ in space-time to an infinite number of degrees of freedom encoded in the wave-function $\psi(x)$ and evolving into a  field $\psi(x,t)$ in space-time. These infinite number of degrees of freedom dress the classical motion, recovered from the expectation values of the position and momentum operators, with extra degrees of freedom given by all the higher moments of the wave-functions. In the semi-classical regime, these describe the shape and fluctuations of wave-packets around the classical trajectory.
%
%Quantum operators acting quantum states become classical differential operators acting the classical field $\psi$.

\medskip

If one is interested in the quadratic moments, putting aside as a first approximation higher moments of the wave-functions, for instance in order to investigate the semi-classical behavior of wave-packets, one can focus on the Gaussian ansatz for the wave-function,
\be
\psi(x,t)=Ne^{i\gamma}\,e^{\f i\hbar p(x-q)}\,e^{-\lambda(x-q)^{2}}
\,,\qquad\textrm{with}\quad
%n=Ne^{i\gamma}\in\C
%\quad \textrm{and}\quad
\lambda=
\f1{4\alpha^{2}}\left(1-\f{2i}\hbar \alpha\beta\right)
\in\C\,,
\ee
where $N(t),\gamma(t),q(t),p(t),\alpha(t),\beta(t)$ are all variables evolving in time.
We fix the real factor $N>0$ by requiring that the wave-function be normalized,
\be
1=\int |\psi|^{2}
=\int\rd x\,N^{2}e^{-\f{(x-q)^{2}}{2\alpha^{2}}}
%=N^{2}\sqrt{\f{\pi}{2\sigma}}
\quad\Rightarrow\quad
N^{2}=\f1{\alpha\sqrt{2\pi}}\,.
\ee
Beside the global phase $\gamma$, the meaning of the other variables $q,p,\alpha,\beta$ is given by the linear and quadratic expectations values in the position and momentum.
This is the kinematics of the Gaussian wave-packets:
\be
\la \hx\ra=q
\,,\qquad
\la \hp\ra=p
\,,\qquad
\la \hx^{2}\ra=q^{2}+\alpha^{2}
\,,\qquad
\la \hp^{2}\ra=p^{2}+\beta^{2}+\f{\hbar^{2}}{4\alpha^{2}}
\,,\qquad
\la \hD\ra=pq+\alpha\beta\,,
\ee
showing that the Gaussian ansatz has the minimal allowed quantum uncertainty:
\be
c=\sigma_{x}\sigma_{p}-\sigma_{D}=\f{\hbar^{2}}4
\,.
\ee

\medskip

Focusing on the Gaussian ansatz reduces the infinite number of degrees of freedom of the wave-function, one d.o.f. per space point, to a finite number of degrees of freedom, given by the free parameters of the ansatz $(\gamma,q,p,\alpha,\beta)$.
Let us insert this ansatz directly in the Schr\"odinger field theory action \eqref{eqn:Schroaction} for quantum mechanics. This is similar to ``mini-superspaces'' used as gravitational models, obtained by injecting a reduced metric ansatz in the Einstein-Hilbert action for general relativity or in more general actions for modified gravity. To have a clear derivation of the resulting effective action for Gaussian wave-packets, as studied previously in \cite{ABCV,Prezhdo2000QuantizedHD,Bojowald:2022lbe,Livine:2023vph}, it is convenient to cut the Schr\"odinger Lagrangian in the kinetic term $L_{kin}$, the Laplacian diffusion term $L_{\Delta}$ and potential term $L_{V}$:
\be
S[\psi]=\int \rd t\,\rd x\,\bigg{[}
\underbrace{{\vphantom{\f17}i\hbar\bpsi\pp_{t}\psi}}_{L_{kin}}
\underbrace{{-\f{\hbar^{2}}{2m}\pp_{x}\bpsi\pp_{x}\psi}}_{L_{\Delta}}
\underbrace{{\vphantom{\f17}-V(x)|\psi|^{2}}}_{L_{V}}
\bigg{]}
\,,
\nn
\ee
Evaluating each of those terms on the Gaussian ansatz yields:
\be
L_{kin} =
p\dot{q}+\beta\dot{\alpha}+\rd_{t}\left[\f12 \alpha\beta -\hbar \gamma\right]\,,
\qquad
L_{\Delta}=-\left[
\f{p^{2}}{2m}+\f{\beta^{2}}{2m}+\f{\hbar^{2}}{8m}\f1{\alpha^{2}}
\right]\,,
\ee
showing that we have two effective canonical pairs of variables, $(q,p)$ and $(\alpha,\beta)$ and that the global phase $\gamma$ is not a free dynamical parameter. The variable $p$ is the momentum conjugate to the mean position $q$, while $\beta$ is the momentum conjugate to the position mean deviation $\alpha$. 
%
%One evaluates the free Laplacian propagation term:
%\beq
%L_{\Delta}&=&
%-\f{\hbar^{2}}{2m}\int\rd x\, N^{2}e^{-2\sigma(x-q)^{2}}
%\left(-\f i\hbar p-2\bar{A}(x-q)\right)\left(\f i\hbar p-2{A}(x-q)\right)
%\\
%&=&
%-\left[
%\f{p^{2}}{2m}+\f{u^{2}}{2m}+\f{\hbar^{2}}{8m}\f1{w^{2}}
%\right]\,,
%\eeq
%reproducing the conformal potential term for 2nd order moments.
%
This extra canonical pair $(\alpha,\beta)$ represents the quadratic spread of the wave-function and effectively plays the role of a kind of ``relative motion'' coupled to the classical ``center-of-mass'' motion of the principal canonical pair $(q,p)$.
In the vanishing potential case, $V=0$, the Laplacian term leads to an effective potential in $\alpha^{-2}$, which exactly matches the effective equations of motion \eqref{eqn:effmotion} and effective Hamiltonian \eqref{eqn:effH} derived earlier in the previous section for a coupling $c=\hbar^{2}/4$.
This shows the consistency and robustness of this effective dynamics for the quantum uncertainty.
Let us insist that this ``quantum mini-superspace'' is simply the truncation of quantum mechanics from the infinite number of degrees of freedom of the wave-function $\psi(x)$ down to a pair of degrees of freedom  representing the position and momentum standard deviations. One could recover a full description of the wave-function by relaxing the Gaussian ansatz and extracting the dynamics of the whole tower of higher moments of the wave-functions, which encode the complete shape of the wave-packet around its mean value (see e.g. \cite{Livine:2023vph}).

\medskip

Finally, one evaluates the potential term:
\be
L_{V}
%=-\int N^{2}Ve^{-\f{(x-q)^{2}}{2\alpha^{2}}}
=
-\f1{\alpha\sqrt{2\pi}}\int Ve^{-\f{(x-q)^{2}}{2\alpha^{2}}}\,,
\ee
\be
V(x)=V_{0}+\f{V_{2}}2x^{2}+\f{V_{4}}{4!}x^{4}+\dots
\quad\Rightarrow\quad
L_{V}
=
-\left[
V(q)+\f{V_{2}}2\alpha^{2}+\f{V_{4}}{8}\alpha^{4}+\f{V_{4}}{4}\alpha^{2}\beta^{2}+\dots
\right]\,,
\ee
where one recognizes the classical potential, plus an effective potential driving the evolution of the quadratic uncertainty, plus terms coupling the quantum uncertainty to the position expectation values, thereby creating a kick-back of the quantum dynamics of the shape of the wave-packet on the classical trajectory. For more applications of this effective quantum mechanics approach for describing and interpreting quantum phenomenology, we refer the interested reader to \cite{ABCV,Prezhdo2000QuantizedHD,PhysRevE.84.026616,Baytas:2018ruy,Baytas:2018gbu}. 

\medskip

This Gaussian ansatz effective formalism fixes the value of the conformal potential coupling to its minimally allowed value $c=\hbar^{2}/4$. 
A natural question is then how can one change the value of this coupling? So, what generalized wave-packet ansatz would thus allow to shift the value of the coupling of the effective quantum potential? And since this coupling has been understood to reflect the uncertainty, and more precisely the Heisenberg inequality, this questions naturally becomes: how can one excite the uncertainty above its minimal allowed value?

%%%%%%%%
\section{The spectrum of quantum uncertainty}
%%%%%%%%

%%%
\subsection{Field observables and uncertainty algebra}
%\subsection{Field observables, uncertainty algebra and second quantization}
%%%

%%%
%\subsection{Uncertainty algebra}
%%%

Let us now consider the evolution of a general wave-function. We totally unfreeze the Gaussian wave-packet ansatz discussed in the previous section and consider the kinematics and dynamics of the quantum uncertainty in full quantum mechanics.

Written in terms of a wave-function $\psi(x)$, the quantum dynamics of the particle, aside from projective measurements, is formulated as a classical  field evolution.
Indeed, starting from Schr\"odinger action principle,
\be
\label{eqn:Schroactionfull}
S[\psi]=\int \rd t\,\rd x\,\left[
i\hbar\bpsi\pp_{t}\psi
-\f{\hbar^{2}}{2m}\pp_{x}\bpsi\pp_{x}\psi
-V(x)|\psi|^{2}
-\sum_{n\ge1}\f{\ka_{n}}n|\psi|^{2(n+1)}
\right]
\,,
\ee
where we have included the possibility non-linear self-interaction terms for the sake of completeness, one can proceed to a Hamiltonian analysis of quantum mechanics as a classical field theory, with a field Poisson bracket,
\be
\label{eqn:psibracket}
\{\psi(x),\bpsi(y)\}=\f1{i\hbar}\delta(x-y)
\,,
\ee
and a field Hamiltonian,
\be
\cH[\psi]
=
\int \rd x\,\left[
\f{\hbar^{2}}{2m}\pp_{x}\bpsi\pp_{x}\psi
+V(x)|\psi|^{2}
+\sum_{n\ge1}\f{\ka_{n}}n|\psi|^{2(n+1)}
\right]
\,.
\ee
This standard field theory formalism naturally deals with non-linear extension of the Schr\"odinger equation. It is the obvious starting point for a second quantization of the theory, to move up from quantum mechanics to quantum field theory. 
Here we will not deal with the QFT dynamics, but will discuss the fate of the quantum uncertainty upon second quantization at the kinematical level.

%Start with non-linear Schr\"odinger equation, with no potential:
%\be
%i\hbar \pp_{t}\psi=-\f{\hbar^{2}}{2m}\pp_{x}^{2}\psi
%+\ka|\psi|^{2}\psi
%\,,
%\ee
%corresponding to field Hamiltonian:
%\be
%\cH=\int \rd x\left[
%\pp_{x}\bpsi\pp_{x}\psi+\f\ka2|\psi|^{4}
%\right]
%\,.
%\ee
%Compute evolution. Number of particles $Q=\int |\psi|^{2}$:
%\be
%i\rd_{t} |\psi|^{2}
%=
%\f{\hbar^{2}}{2m}\pp(\psi\pp\bpsi-\bpsi\pp\psi)
%\,,\qquad
%\rd_{t}Q=\rd_{t}\int |\psi|^{2}=0
%\,.
%\ee
%Defines constant of motion, interpreted through Noether's theorem as generator of symmetry under change of global phase.
%%
%Current:
%\be
%i\hbar\rd_{t}(\psi\pp\bpsi-\bpsi\pp\psi)
%=
%\pp(\psi\pp^{2}\bpsi+\bpsi\pp^{2}\psi-2\pp\psi\pp\bpsi)-\ka\pp|\psi|^{4}
%\,.
%\ee
%Compute trivially, evolution of momentum:
%\be
%\rd_{t}\la p \ra =0
%\,.
%\ee
%%
%Corresponds from translation-invariance of theory, even with self-interaction.
%%
%Position:
%\be
%\la x\ra=\int x|\psi|^{2}
%\,,\qquad
%\rd_{t}\la x\ra =\f{i\hbar}{2m}\int \psi\pp\bpsi-\bpsi\pp\psi
%=
%\f1m\la p \ra
%\,.
%\ee
%Defines Galilean charge $m\la x \ra-t\la p\ra$, explicitly time dependent, and generates symmetry under Galilean boosts.
%%
%Uncertainty:
%\be
%\rd_{t}\la x^{2}\ra
%=
%\f2m\la xp\ra_{sym}\,
%\ee
%\be
%\rd_{t}\la xp\ra_{sym}
%=
%\f1m\la p^{2}\ra
%+\f\ka 2
%\int|\psi|^{4}
%\,,
%\ee
%\be
%i\hbar \rd_{t}
%\int|\psi|^{4}
%=
%\f{\hbar^{2}}m
%\int \bpsi^{2}(\pp\psi)^{2}-\psi^{2}(\pp\bpsi)^{2}
%\,\dots
%\ee
%Could push further. In fact, known that integrable structure (cite Russian paper). 
%

Expectation values of quantum operators, correlations functions and self-interaction terms are then mathematically all on the same levels and defined as classical field observables. So we define:
\be
Q=\int \rd x\, |\psi|^{2}
\,,\quad
\la x \ra=\int \rd x\, x|\psi|^{2}
\,,\quad
\la p \ra=-i\hbar\int \rd x\, \bpsi\pp_{x}\psi
\,,
\ee
\be
\la x^{2} \ra=\int \rd x\, x^{2}|\psi|^{2}
\,,\quad
\la p^{2} \ra=-\hbar^{2}\int \rd x\, \bpsi\pp_{x}^{2}\psi
\,,\quad
\la D \ra=-\f {i\hbar}2\int \rd x\, \bpsi\,\big{[}x\pp_{x}+\pp_{x}x\big{]}\psi
\,,
\ee
The number of particles $Q$, or wave-function normalization, is the generator of phase transformations on the wave-function, and Poisson-commutes with the linear and quadratic moments,
\be
\{Q,\la x\ra\}=\{Q,\la p\ra\}=\{Q,\la x^{2}\ra\}=\{Q,\la p^{2}\ra\}=\{Q,\la D\ra\}=0
\,.
\ee
The linear expectation values have a canonical Poisson bracket,
\be
\{\la x \ra,\la p\ra\}=Q
\,,
\ee
while the quadratic moments' Poisson brackets form a closed Lie  algebra:
\be
\{\la x^{2}\ra,\la p^{2}\ra\}=4\la D\ra
\,,\qquad
\{\la D\ra,\la x^{2}\ra\}=-2\la x^{2}\ra
\,,\qquad
\{\la D\ra,\la p^{2}\ra\}=2\la p^{2}\ra
\,.
\ee
This echoes the  $\sl(2,\R)$ operator algebra formed by the quadratic polynomials $\hx^{2}$, $\hp^{2}$ and $\hD$, as written in \eqref{eqn:sl2}. The $\sl(2,\R)$ Casimir of this algebra of field observables is:
\be
\cC=\la x^{2}\ra\la p^{2}\ra-\la D \ra^{2}
\ee
Heisenberg uncertainty relation, or more precisely Robertson-Schr\"odinger inequality, implies that
\be
\cC\ge Q^{2} \f{\hbar^{2}}4 >0
\,,
\ee
where we made sure to keep track of the normalization $Q$ factor of the wave-function.
In order to focus on the uncertainty and quantum fluctuations, we introduce the standard deviation observables:
%
%Actually can introduce ``reduced'' $\sl_{2}(\R)$ algebra for quantum uncertainty fluctuations:
\be
\varsigma_{x}=Q\la x^{2}\ra-\la x \ra^{2}
\,,\qquad
\varsigma_{p}=Q\la p^{2}\ra-\la p \ra^{2}
\,,\qquad
\varsigma_{D}=Q\la D\ra-\la x \ra \la p\ra\,,
\ee
where we adjust the normalization factors to account that the wave-function is not necessarily normalized to 1.
Computing the Poisson brackets of these observables\footnotemark{},
\footnotetext{
As an intermediate step, we point out the following Poisson brackets:
\be
\{\la x^{2} \ra,\la p \ra\}=2\la x\ra\,,\quad\{\la x \ra,\la p^{2} \ra\}=2\la p \ra\,,\quad
\{D,\la x \ra\}=-\la x \ra\,,\quad\{D,\la p \ra\}=+\la p \ra\,.
\nn
\ee
}
we obtain a ``reduced'' $\sl(2,\R)$ algebra for quantum uncertainty fluctuations: 
\be
\{\varsigma_{x},\varsigma_{p}\}
%=
%4Q^{2}D+4Q\la x \ra\la p \ra -4Q\la x\ra\la p \ra-4Q\la x\ra\la p \ra
=
4Q\varsigma_{D}
\,,\qquad
\{\varsigma_{D},\varsigma_{p}\}
%=
%2Q^{2}\la p^{2}\ra-2Q\la p\ra^{2}-2Q\la p\ra^{2}+2Q\la p\ra^{2}
=
2Q\varsigma_{p}
\,,\qquad
\{\varsigma_{D},\varsigma_{x}\}
=
-2Q\varsigma_{x}
\,.
\ee
The number observable $Q$ commutes with all the observables, and it should thus be considered as a central charge.
We then introduce the reduced $\sl(2,\R)$ Casimir, which measures the quantum uncertainty:
\be
c=\varsigma_{x}\varsigma_{p}-\varsigma_{D}^{2}\,.
\ee
Robertson-Schr\"odinger  inequality for the quantum uncertainty ensures that $c\ge Q^{4}\hbar^{2}/4>0$.

%\medskip

%%%
\subsection{Second quantization and wave-function correlation functionals}
\label{sec:QFT}
%%%

Let us now canonically quantize the wave-function as a field. We introduce the space of fields $\psi$. We will assume $\psi$ to be rapidly decreasing, and we would like to consider wave-functionals $\Psi[\psi]$.
This wave-functional $\Psi$ is to be thought as a wave-function of the wave-function.
%
%, living in the closure of polynomials in the field $\psi$.
%
We will focus here on polynomial holomorphic wave-functionals.

We raise the canonical bracket \eqref{eqn:psibracket} between the wave-function and its complex conjugate to operators acting on wave-functionals,
%the space of fields,
\be
\psi(x)\rightarrow \widehat{\psi(x)}=\psi(x)\times
\,,\qquad
\bpsi(x)\rightarrow \widehat{\bpsi(x)}=\f{\delta}{\delta\psi(x)}
\,.
\ee
The operators $\widehat{\psi(x)}$ play the role of annihilation operators, while the operators $\widehat{\bpsi(x)}$  play the role of creation operators and act as differential operators on the space of holomorphic functionals.
%
%The study here is of restricted scope, we will not carefully define the space of functional, the measure over the fields and scalar product between functionals.
%
We will not describe here the full second quantization, i.e. the identification and quantization of a complete set of field-observables.
We would like to focus instead on the quantization of the field observables representing the quadratic moments of the wave-functions as operators acting on functionals $\Psi[\psi]$. 

We proceed to the straightforward quantization of the moments of the wave-function, using a normal ordering,
\be
\hQ=\int\rd x\,\psi(x)\f{\delta}{\delta\psi(x)}
\,,\quad
\hX\equiv\widehat{\la x \ra}
=\int\rd x\,x\psi(x)\f{\delta}{\delta\psi(x)}
\,,\quad
\hP\equiv\widehat{\la p \ra}
=-i\hbar\int\rd x\,\pp_{x}\psi(x)\f{\delta}{\delta\psi(x)}
\,,
\ee
\be
\hK_{D}\equiv\widehat{\la D \ra}=-{i\hbar}\int \rd x\,
\Big{[}x\pp_{x}\psi+\psi(x)\Big{]}\,\f{\delta}{\delta\psi(x)}
\,,
\ee
\be
\hK_{x}\equiv\widehat{\la x^{2} \ra}=\int \rd x\,x^{2}\psi(x)\f{\delta}{\delta\psi(x)}
\,,\quad
\hK_{p}\equiv\widehat{\la p^{2} \ra}=-\hbar^{2}\int \rd x\,\pp^{2}\psi(x)\f{\delta}{\delta\psi(x)}\,,
\ee
where we renamed the field operators in order to keep  notations as simple as possible.
This definition ensures that the quadratic moment operators preserve the $\sl(2,\R)$ Lie algebra structure
\be
[\hK_{x},\hK_{p}]=-4i\hbar \hK_{D}
\,,\quad
[\hK_{D},\hK_{x}]=+ 2i\hbar \hK_{x}
\,,\quad
[\hK_{D},\hK_{p}]=- 2i\hbar \hK_{p}
\,,\quad
[\hQ,\hK_{\mu}]=0
\,.
\ee
Here we have chosen by hand a particular operator ordering, in order to preserve this algebraic structure. The choice of ordering should more rigorously be fixed by requiring to have self-adjoint operators, once a scalar product on the space of functionals is defined. Since the operators considered above are of first order differential, changing the ordering can only produce constant shifts and will not affect the algebraic and analytical properties of those operators.

\medskip

We would like to study the action of those operators on polynomial weakly-continuous wave-functionals $\Psi[\psi]$. More precisely, we will consider $N$-point correlation functionals defined as,
\be
\Psi^{(N)}_{\phi^{(N)}}[\psi]\equiv \int\prod_{k=1}^{N}\rd x_{k}\, \phi^{(N)}(x_{1},..,x_{N})\prod_{k}^{N}\psi(x_{k})
%\,,\qquad
%\hQ\,\Psi^{(2)}_{\phi}=2\,\Psi^{(2)}_{\phi}
\,,
\ee
where we will assume, to be on the safe side of integrations and functional derivations, that the wave-function $\psi$ is rapidly decreasing and that the test functions $\phi^{(N)}$ are smooth functions on $\R^{N}$.
%
%Rigorously building a Hilbert space of wave-functionals is a very subtle question in functional analysis. The study here is of restricted scope. We will not attempt to define the measure over fields. Although we clearly have in mind a scalar product between functionals as an infinite product integration over all Fourier modes of the field, we will make use of such a formula. We will nevertheless make the implicit reasonable assumption that the polynomial wave-functionals defined above do belong to a well-defined Hilbert space of wave-functionals and that the quadratic moment operators are self-adjoint differential operators acting on that Hilbert space. 
%
%Although this will definitely need to be checked in order to set the present analysis on solid mathematical grounds, one can consider the fact that we do recover the expected spectrum for the $\sl(2,\R)$ Casimir operator for unitary representations, as we show below, as a convincing hint that this assumption does hold.
%
These correlations functionals are actually standard $N$-particle states of quantum field theory, usually written in the momentum basis but here defined in real space,
\be
\Psi^{(N)}_{\phi^{(N)}}=
\int \prod_{k=1}^{N}\f{\rd p_{k}}{2\pi}\,
\phi^{(N)}_{p_{1},..,p_{N}}
\,|p_{1},..,p_{N}\ra
\,,\quad\textrm{with}\quad
\phi^{(N)}_{p_{1},..,p_{N}}
=
\prod_{k=1}^{N}\rd x_{k}\, \phi^{(N)}(x_{1},..,x_{N})e^{i\sum_{k}p_{k}x_{k}}
\,,
\ee
so that one can use the standard definition of quantum field states used in quantum field theory.

As indicated by its name, the number of particle operator $\hQ$ is diagonal on $N$-point correlation functionals $\Psi^{(N)}$. Its spectrum is the set of integers $\N$. It simply probes the power of the polynomial and gives the number of particles $N$.
%
%Starting with $\hQ$, it is the number of particle operator. It sees the power of the polynomial of the functional $\Psi[\psi]$ in the wave-function $\psi$, and its spectrum is the set of integers $\N$.
%
For instance, the linear wave-functionals in $\psi$ are the one-particule states,
\be
\Psi^{(1)}_{\vphi}[\psi]\equiv \int\rd x\, \vphi(x)\psi(x)
\,,\qquad
\hQ\,\Psi^{(1)}_{\vphi}=\Psi^{(1)}_{\vphi}
\,,
\ee
and are smooth smearings of the wave-function over space.
Two-particle states are the quadratic functionals,
\be
\Psi^{(2)}_{\phi}[\psi]\equiv \int\rd x\,\rd y\, \phi(x,y)\psi(x)\psi(y)
\,,\qquad
\hQ\,\Psi^{(2)}_{\phi}=2\,\Psi^{(2)}_{\phi}
\,,
\ee
and probe the 2-point correlations of the wave-function. These contain the same information as the Gaussian functionals usually considered in quantum field theory,
\be
\cG_{\sigma(p)}[\psi]=e^{-\int\rd p\,\sigma(p)\psi_{-p}\psi_{+p}}\,,
\quad\textrm{with}\quad
\psi_{p}=\int \rd x \, \psi(x)e^{-i px}\,,
\ee
parametrized by an arbitrary Gaussian width function $\sigma(p)$ and defined in terms of the Fourier components $\psi_{p}$ of the wave-function.
%\be
%\psi_{p}=\int \rd x \, \psi(x)e^{-i px}
%\,.
%\ee
This type of dependency on the wave-function is reproduced by considering 2-point correlations with a test field $\phi$ depending only on the position difference $(x-y)$:
\be
\phi(x,y)=\int \f{\rd p}{2\pi}\,\sigma(p)e^{-ip(x-y)}
\quad\Rightarrow\quad
\Psi^{(2)}_{\phi}[\psi]= \int\rd x\,\rd y\, \phi(x,y)\psi(x)\psi(y)
=
\int\rd p\,\sigma(p)\psi_{-p}\psi_{+p}
\,.
\ee
Then, higher correlation functionals for $N\ge 3$ correspond to states with higher number of particles and allow to represent non-Gaussianities in quantum field theory.

\medskip

Now, let us turn to the  $\sl(2,\R)$ Lie algebra generated by $\hK_{x}=\widehat{\la x^{2}\ra}$, $\hK_{p}=\widehat{\la p^{2}\ra}$ and $\hK_{D}=\widehat{\la D\ra}$. Since we would like keep those operators Hermitian, we are interested in unitary representations of $\sl(2,\R)$. Irreducible unitary representations of $\SL(2,\R)$ are well-known (see \cite{lang2011sl2,knapp2016representation} for textbooks, see also e.g. \cite{Kitaev:2017hnr,Freidel:2002hx,Conrady:2010sx}) and can be classified according the value of the Casimir operator,
\be
\hat{\cC}=\f12(\hK_{p}\hK_{x}+\hK_{x}\hK_{p})-\hK_{D}^{2}
\,,
\ee
We distinguish three classes of representations:
\begin{itemize}
\item The positive and negative discrete series of unitary representation labeled by a half-integer $j\ge 0$ with Casimir values $4\hbar^{2}j(j+1)$. These are lowest and highest weight infinite-dimensional irreducible unitary representations of $\sl(2,\R)$, meaning that the eigenvalues for the operator $\hK_{x}+\hK_{p}$ are  either bounded from below by $+j$ (positive series) or bounded from above by $-j$ (negative series).

\item The principal continuous series of irreducible unitary representations labeled by a real number $s\in\R$ with Casimir values $s^{2}+1$ always larger than 1.

\item The complementary (or exceptional) continuous series of irreducible unitary representations with Casimir values running in the real interval $]0,1[$.

\end{itemize}
Since the uncertainty is ``classically'' positive in quantum mechanics, $\cC\ge \hbar^{2}/4>0$, due to the Heisenberg inequality, we expect that the discrete series of representations with positive Casimir will be the relevant ones realized here, so that we expect a spectrum of $\hat{\cC}$ in $4j(j+1)$ for half-integers $j=\in\N/2$.
% plus possibly the limit case representation $j=-\f12$.
%
We are not going to investigate the whole spectrum in a rigorous manner, but focus instead on the $\sl(2,\R)$ action on one-particle and two-particle states.

We compute the action of the uncertainty generators $\hK_{\mu}$ on one-particle functionals,
\be
\hK_{x}\Psi^{(1)}_{\vphi}=\Psi^{(1)}_{x^{2}\vphi}
\,,\qquad
\hK_{p}\Psi^{(1)}_{\vphi}=-\hbar^{2}\Psi^{(1)}_{\pp_{x}^{2}\vphi}
\,,\qquad
\hK_{D}\Psi^{(1)}_{\vphi}=i\hbar\Psi^{(1)}_{x\pp_{x}\vphi+\f12\vphi}
\,,
\ee
which allows to compute the Casimir of the $\sl_{2}$ uncertainty algebra:
\be
\hat{\cC}\Psi^{(1)}_{\vphi}=-\hbar^{2}\f34\Psi^{(1)}_{\vphi}\,.
\ee
Comparing this value against the classification reviewed above, this corresponds to a representation from the exceptional continuous series. 
This does not match the minimal value of the uncertainty allowed in standard quantum mechanics, $\cC_{min}=\hbar^{2}/4$. This spectrum shift
%due to quantization
is analogous to the shift in the harmonic oscillator energy spectrum due to the operator non-commutativity. One can thus wonder if this uncertainty shift has any meaningful physical consequence.  Putting it in light in experiments would illustrate the difference between quantum mechanics and quantum field theory, and would underline the necessity of the ``2nd quantization'' even in the non-relativistic regime.

Moving up to the 2-particle sector, we similarly compute the action of the $\hK_{\mu}$ operators,
\be
\hK_{x}\Psi^{(2)}_{\phi}=\Psi^{(2)}_{(x^{2}+y^{2})\phi}
\,,\qquad
\hK_{p}\Psi^{(2)}_{\phi}=-\hbar^{2}\Psi^{(2)}_{\pp_{x}^{2}\phi+\pp_{y}^{2}\phi}
\,,\qquad
\hK_{D}\Psi^{(2)}_{\phi}=i\hbar\Psi^{(2)}_{x\pp_{x}\phi+y\pp_{y}\phi+\phi}
\,,
\ee
which leads to a very elegant formula for the $\sl_{2}$ Casimir in terms of an angular momentum operator,
\be
\hat{\cC}\Psi^{(2)}_{\phi}=\Psi^{(2)}_{\hat{\cC}\triangleright\phi}
\qquad\textrm{with}\quad
\hat{\cC}\triangleright\phi=
\hbar^{2}\Big{[}
\big{(}i(x\pp_{y}-y\pp_{x})\big{)}^{2}-1
\Big{]}\, \phi
\,,
\ee
so that its spectrum is obviously labeled by an integer $n\in\N$,
\be
\hat{\cC}\Psi^{(2)}_{\phi}
=
\hbar^{2}\big{(}
n^{2}-1
\big{)}
\Psi^{(2)}_{\phi}
=
4\hbar^{2}\,j(j+1)
\Psi^{(2)}_{\phi}
\qquad\textrm{with}\quad
n=2j+1\,.
\ee
Thus the 2-particle sector allows us to recover the whole expected spectrum with all the discrete series of $\sl(2,\R)$ representations.

For instance, for $\phi=1$, corresponding to the functional $\Psi_{1}^{(2)}[\psi]=(\int \psi)^{2}$, we get the minimal value $n=0$, corresponding to the null limit case of the discrete series of representations. Similarly, the choice $\phi=(x^{2}+y^{2})$ is invariant under rotations in the $(x,y)$ plane, so that the corresponding functional $\Psi^{(2)}_{x^{2}+y^{2}}[\psi]\propto(\int x^{2}\psi)\int \psi$ also gives the  $n=0$ eigenvalue.

A first non-trivial case is $\phi=xy$, which still corresponds to a factorizable correlation functional $\Psi^{(2)}[\psi]\propto(\int x\psi)^{2}$, gives the eigenvalue $n=2$. More generally, one can identify a whole set of uncertainty Casimir  eigenstates:
\be
\phi_{n}(x,y)=e^{in\arctan\f yx}
\quad\Rightarrow\quad
\hat{\cC}\Psi^{(2)}_{\phi_{n}}
=
\hbar^{2}\big{(}
n^{2}-1
\big{)}
\Psi^{(2)}_{\phi_{n}}
\,.
\ee
These eigenstates define specific 2-point correlation functionals, which are simple to write in polar coordinates in the $(x,y)$ plane,
\be
\Psi^{(2)}_{\phi_{n}}[\psi]
=
\int\rd x\rd y\,e^{in\arctan\f yx}\,\psi(x)\psi(y)
=
\int r \rd r\rd\theta\,
e^{in\theta}\psi(r\cos\theta)\psi(r\sin\theta)
\,.
\ee
This expression nevertheless remains mysterious considering that we are studying 1d mechanics and that $x$ and $y$ are coordinates on the same spatial axis. It would be interesting to further understand what properties of the wave-functions are probed and revealed by these correlation integrals.

\medskip

We can similarly study the ``reduced'' $\sl(2,\R)$ algebra generated by the standard deviation observables $\varsigma_{\mu}$ once quantized,
\be
\hvsigma_{x}=\hQ\hK_{x}-\hX^{2}
\,,\qquad
\hvsigma_{p}=\hQ\hK_{p}-\hP^{2}
\,,\qquad
\hvsigma_{D}=\hQ\hK_{D}-\f12(\hX\hP+\hP\hX)
\,,
\ee
where we remember that the operator $\hQ$ commutes with all the other considered operators. Straightforward calculations\footnotemark{} allow to check that the reduced uncertainty Casimir $\hc$ vanishes on 1-particle functionals and is constant on the 2-particle sector,
\footnotetext{
We compute the action of the $\hvsigma_{\mu}$ operators on the 2-point correlations:
\be
\hvsigma_{x}\Psi^{(2)}_{\phi}
=
\Psi^{(2)}_{(x-y)^{2}\phi}
\,,\qquad
\hvsigma_{p}\Psi^{(2)}_{\phi}
=
-\hbar^{2}\Psi^{(2)}_{(\pp_{x}-\pp_{y})^{2}\phi}
\,,\qquad
\hvsigma_{D}\Psi^{(2)}_{\phi}
=
i\hbar\Psi^{(2)}_{(x-y)(\pp_{x}-\pp_{y})\phi+\phi}
\,.
\nn
\ee
}
\be
\hc=\f12(\hvsigma_{x}\hvsigma_{p}+\hvsigma_{p}\hvsigma_{x})-\hvsigma_{D}^{2}
\,\qquad
\hc\Psi^{(1)}_{\vphi}=0
\,,\quad
\hc\Psi^{(2)}_{\phi}=-2^{2}\hbar^{2}\,\f34\Psi^{(2)}_{\phi}\,.
\ee
Keeping in mind the $Q^{2}$ factor in front of the Casimir, here for $Q=2$, the $\hc$ eigenvalue on 2-point correlations clearly corresponds to the $\hC$ eigenvalue on 1-point correlations. This hints to the possibility that the reduced uncertainty operators $\hvsigma_{\mu}$ simply reproduce the quadratic uncertainty operators $\hK_{\mu}$ pushed to one-point higher correlations. This makes sense from the persepctive that the standard deviation observables, $\varsigma_{x},\varsigma_{p},\varsigma_{D}$, are simply the quadratic moments, $\la x^{2}\ra$, $\la p^{2}\ra$, $\la D\ra$, once setting the averages to 0.

To substantiate this claim, we push our analysis to the 3-particle sector, introducing the 3-point correlations,
\be
\Psi^{(3)}_{f}[\psi]
=\int\rd x \,\rd y\,\rd z\, f(x,y,z)\psi(x)\psi(y)\psi(z)
\,.
\ee
We similarly compute the action of all field operators on these functionals and obtain, after a slightly tedious and non-enlightening calculation,
\be
\hc\Psi^{(3)}_{f}
=\Psi^{(3)}_{\hc\triangleright f}
\qquad\textrm{with}\quad
\hc\triangleright f
=
3^{2}\hbar^{2}\Bigg{[}
\f13\Big{(}i(
x\pp_{y}-y\pp_{x}+y\pp_{z}-z\pp_{y}+z\pp_{x}-x\pp_{z}
)\Big{)}^{2}
-1
\Bigg{]}
\,.
\ee
We recognize an angular momentum differential operator,
\be
i(
x\pp_{y}-y\pp_{x}+y\pp_{z}-z\pp_{y}+z\pp_{x}-x\pp_{z}
)
=J_{1}+J_{2}+J_{3}\,,
\ee
with the standard 3d notations, $J^{a}=i\eps^{abc} x_{b}\pp_{c}$. Since the vector $(1,1,1)\in\R^{3}$  is of squared-norm 3, we recover the expected spectrum labeled by integers $n\in\N$,
\be
\hc\Psi^{(3)}_{f}=
3^{2}\hbar^{2}\,\big{(}n^{2}-1\big{)}\,\Psi^{(3)}_{f}
\,,
\ee
where we recognize the 2-particle spectrum of the $\sl_{2}$ Casimir $\hC$ up to the extra factor $Q^{2}$, here for $Q=3$. Writing the eigenmodes of the angular momentum operator, we could write explicitly the corresponding 3-point correlations, but that didn't lead to any insight beyond the discussion above for 2-point correlations.

\medskip

To summarize this section, pushing the logic that quantum mechanics is to be upgraded to quantum field theory, we have promoted the wave-function $\psi$  to a wave-functional $\Psi[\psi]$. To avoid awkward nomenclature, we have used standard quantum field theory terminology and referred to the quantum mechanics' wave function $\psi$ as a field and considered wave-functionals over the space of fields in the context of a 2nd quantization. We have thus proceeded to the quantization of field observables, focusing on representing the quadratic moments and standard deviations of the wave-function. These quadratic moments form a $\sl_{2}$ Lie algebra, preserved at the quantum field theory level, and the $\sl_{2}$ Casimir quantifies the quantum uncertainty from Heisenberg uncertainty inequality. Finally, we have derived a discrete spectrum from the quantum uncertainty in this second quantization of quantum mechanics.

This leads for instance to a shift between the minimal value of the uncertainty in quantum mechanics from Heisenberg inequality and the minimal eigenvalue for the quantum uncertainty operator after 2nd quantization. This kind of effects might possibly translate into interesting phenomenology and lab experiments investigating the physical signatures of 2nd quantization and the the differences between quantum mechanics (QM) and quantum field theory (QFT) in the non-relativistic regime.

However, working with wave-functionals of the wave-function does not look very enlightening and clearly leads to a deeper ontological questioning on the meaning of quantum states and the physical meaning of quantizing a (field) theory. Indeed, although showing that the uncertainty operator acting on
%2-particle quantum fields
2-point correlations of the wave-function
is expressed as an angular momentum operator seems to be an elegant mathematical result, it does not explain how one could prepare arbitrary $N$-particle quantum field wave-functions $\Psi^{(N)}[\psi]$.
To clarify the physical meaning of the discrete spectrum of the quantum uncertainty, we thus move one step down, coming back from the 2nd quantized theory to the original quantum mechanics framework. And we look for the equivalent of the closed orbits for the Born-Sommerfeld semi-classical quantization, responsible for this discrete spectrum of the uncertainty: can one identify wave-functions whose  uncertainty matches the eigenvalues of the quantum uncertainty operator?

Answering this question is the purpose of the next section.

%%%%%%%%
\section{Exciting the uncertainty}
%\section{The spectrum of quantum uncertainty}
%%%%%%%%

%%%
\subsection{Proper modes of quantum uncertainty}
%\subsection{Uncertainty excitation}
%\subsection{Uncertainty excitation and extended Gaussian wave-packets}
%%%

Identifying  the wave-functions with non-minimal values of the quantum uncertainty certainly is a simpler problem than diagonalizing the quantum uncertainty operator on the space of wave-function functionals.
Starting from the Gaussian ansatz, already discussed in section \ref{sec:Gaussian}, we construct wave-functions with an excited quantum uncertainty, as when working with an harmonic oscillator.
This leads to the wave-function ansatz, 
\be
\label{eqn:extendedGaussian}
\psi_{k,q,A,\gamma,\cN}(x)=
\cN e^{i\gamma}H_{k}\Big{[}(x-q)/\alpha\sqrt{2}\Big{]}\,e^{\f i\hbar p (x-q)}e^{-\lambda(x-q)^{2}}\,,
\qquad\textrm{with}\quad
\lambda=\f{2k+1}{4\alpha^{2}}-\f{i}{2\hbar} \f\beta\alpha
%\lambda=\f1{4\alpha^{2}}\left(1-\f{2i}\hbar \alpha\beta\right)
\,,
\ee
where the $H_{k}$'s with $k\in\N$ are the Hermite polynomials defined as orthogonal polynomials,
\be
\int_{\R} \rd z\,H_{k}[z]H_{l}[z]e^{-z^{2}}
=\delta_{kl}\sqrt{\pi}2^{k}k!
\,,
\ee
or equivalently constructed from the following recursion relations, with initial conditions $H_{0}[z]=1$ and $H_{1}[z]=2z$,
\be
%H_{0}[z]=1
%\,,\quad
%H_{1}[z]=2z
%\,,\qquad
zH_{k}=\f12H_{k+1}+kH_{k-1}\,,\qquad
H'_{k}=2zH_{k}-H_{k+1}=2kH_{k-1}
\,.
\ee
This is actually the standard extended Gaussian ansatz used to study squeezed wave-packets, e.g. \cite{Zelaya:2021vxv,Gazeau:2022pzz}.
The parameters $(q,p)$ ares respectively the mean position of the wave-packet and momentum carried by the wave-packet. The  parameter $\lambda\in\C$ is the complex width of the Gaussian , $\cN>0$ is a normalization factor, $\gamma\in\R$ is the global phase, and the integer $k\in\N$ is the level of the Hermite polynomial.
The squared norm of the state is
\be
Q=\int |\psi|^{2}=\cN^{2}\,2^{k}k!\,\alpha\sqrt{\f{2\pi}{2k+1}}
\,.
\ee
Imposing $Q=1$ to work with a normalized wave-function, we compute the linear and quadratic expectation values of this ansatz,
\be
\la x\ra=q
\,,\quad
\la p \ra=p
\,,
\ee
\be
\la x^{2} \ra=q^{2}+\alpha^{2}
\,,\qquad
\la p^{2}\ra=p^{2}+\beta^{2}+\f{(2k+1)^{2}\hbar^{2}}{4\alpha^{2}}
\,,\qquad
\la D\ra=pq+\alpha\beta
\,,
\ee
leading to a quantum uncertainty
\be
c
=
\Big{(}\la x^{2} \ra-\la x\ra^{2}\Big{)}\Big{(}\la p^{2} \ra-\la p\ra^{2}\Big{)}
-
\Big{(}\la D\ra-\la x\ra\la p\ra\Big{)}^{2}
%=
%\f{2k+1}{4\sigma}\hbar^{2}\f{2k+1}{\sigma}(\sigma^{2}+\alpha^{2})
%-\left(\hbar\f{2k+1}{2\sigma}\alpha\right)^{2}
=(2k+1)^{2}\f{\hbar^{2}}4
=\left(k+\f12\right)^{2}\hbar^{2}
\,,
\ee
which almost reproduces  the exact quantum spectrum $(n^{2}-1)\hbar^{2}$ with $n\in\N$ given by the $\sl_{2}$ Casimir for the discrete positive series of $\sl(2,\R)$ unitary representations, as derived in the previous section. We indeed recover the correct scaling and leading order  in the integer label, but differ in the $+\f12$ and $-1$ shifts.

The uncertainty level $k$ can be changed and excited, without changing the number of particle, by acting  with the multiplication by $z$ or the differentiation $\pp_{z}$ in terms of the rescaled position $z\equiv (x-q)/\alpha\sqrt{2}$.  It is similar to a notion of intrinsic quantum energy. In the case of a harmonic oscillator, the energy is indeed exactly the uncertainty level.

\medskip

%Fixing $k$ somewhat corresponds to the proper modes of the quantum uncertainty and is the analogue of the quantized radius of Bohr-Sommerfeld orbits.
%
One can actually consider an arbitrary polynomial factor in front of the pure Gaussian and get a superposition of these eigenmodes of the quantum uncertainty. This produces arbitrary real values of $c$ (always larger than the minimal value $\hbar^{2}/4$), so that there is no actual quantization of the uncertainty in quantum mechanics, just as orbits in the 2-body problem can have arbitrary radius in classical mechanics and are only assumed to take discrete values to match the spectrum from quantum mechanics. So it is truly the 2nd quantization of the wave-function which is responsible for the discrete spectrum of the quantum uncertainty.
%

%%%
%\subsection{Exciting the quantum uncertainty}
%%%

%{\bf Exciting the quantum uncertainty. Operators which allow to change uncertainty level $k$}
%
%Using recursion relations on Hermite polynomial, operator as differntial operators in rescaled position $z\equiv \sqrt{2\sigma}(x-q)$:
%\be
%\left(1-\f A{\sqrt{2\sigma}}\right)2z-\f1{\sqrt{2\sigma}}\pp_{z}\dots
%\ee
%
%Better say that multiplication by x and derivation $\pp_{x}$ lead to superposition of higher and lower quantum uncertainty.

%%%%%%%%
%\section{Effective dynamics}
%%%%%%%%

%%%
\subsection{Effective dynamics for extended Gaussian wave-packets}
%%%

The effect of working with  extended Gaussian wave-packets with excited uncertainty is well visualized in the framework of effective dynamics of wave-packets. Repeating the steps of the earlier analysis done for the pure Gaussian ansatz in section \ref{sec:Gaussian}, we show below that the uncertainty excitation level $k$ enters directly the coupling of the effective conformal potential driving the evolution of the position spread.

There are two ways to proceed to derive the effective dynamics of wave-packets. Once  a wave-packet ansatz is  chosen, we can follow two paths:
\begin{itemize}
\item Either one inserts the ansatz in the Schr\"odinger equation, derives the induced equations of motion for the ansatz's parameters and realizes that they can be formulated in terms of an effective Hamiltonian dynamics;
\item Or one inserts the ansatz directly in the Schr\"odinger field action, compute the effective Lagrangian which now depend on the  ansatz's parameters an their time derivative and perform the Legendre transform to obtain an effective Poisson bracket and Hamiltonian.
\end{itemize}
In the present case of the free Schr\"odinger equation, with vanishing potential (and no self-interaction term), it turns out that both methods applied to the extended Gaussian ansatz lead the exact same result.

\medskip

Let us thus start with plugging the extended Gaussian  ansatz \eqref{eqn:extendedGaussian} in the free Schr\"odinger equation,
\be
\pp_{t}\psi=\f{i\hbar}{2m}\pp_{x}^{2}\psi
\,,
\ee
gives equations of motion for the Gaussian parameters. The evolution of the position, momentum, normalization factor and phase are as expected:
\be
\dot{q}=\f pm\,,\quad
\dot{p}=0\,,\quad
\dot{Q}=0\,,\quad
\dot{\gamma}=\f1\hbar\left[
\f{p^{2}}{2m}-(2k+1)\f{\hbar^{2}}{4m}\f1{w^{2}}
\right]\,.
\ee
We further get a Ricatti equation for the complex Gaussian width, which is easily translated into equations on the position and momentum spreads:
\be
\dot{\lambda}=-\f{2i\hbar}{m}\lambda^{2}
\qquad\Rightarrow\quad
\dot{\alpha}=\f{\beta}m\,,\quad
\dot{\beta}={(2k+1)^{2}}\f{\hbar^{2}}{4m}\f1{\alpha^{3}}\,.
\ee
%
%\be
%\left[
%\f{\dot{n}}{n}
%-\f i\hbar p\dot{q}
%+\f i\hbar\dot{p}(x-q)
%+2A\dot{q}(x-q)
%-\dot{A}(x-q)^{2}
%\right]
%\,H_{k}
%+\left[
%\f{\dot{\sigma}}{\sqrt{2\sigma}}(x-q)
%-\dot{q}\sqrt{2\sigma}
%\right]\,H'_{k}
%\ee
%\be
%=
%\f{i\hbar}{2m}
%\left[
%2\sigma H''_{k}
%-2AH_{k}
%+2\sqrt{2\sigma}\left(\f {ip}\hbar-2A(x-q)\right)H'_{k}
%+\left(\f {ip}\hbar-2A(x-q)\right)^{2}H_{k}
%\right]
%\ee
%Introduce:
%\be
%n=Ne^{i\gamma}\,,\quad
%Q\propto\f{N^{2}}{\sqrt{\sigma}}
%\ee
%\be
%w^{2}=\f{2k+1}{4\sigma}\,,\quad
%A=\sigma+i\alpha
%=\f{2k+1}{4w^{2}}-i\f{u}{2\hbar w}
%\,.
%\ee
%such that:
%\be
%\la D \ra =pq+uw\,,
%\qquad
%\la x^{2}\ra=q^{2}+w^{2}
%\,.
%\ee
%
As was previously reviewed in section \ref{sec:evolution} for the pure Gaussian ansatz, these evolution equations for $\alpha$ and $\beta$ can be derived from an effective Poisson bracket and effective Hamiltonian,
\be
\label{eqn:effHk}
\{\alpha,\beta\}_{eff}=1
\,,\quad
H^{(2)}_{eff}=\f{\beta^{2}}{2m}+\f{c}{2m\alpha^{2}}
\qquad\textrm{with}\quad
c=(2k+1)^{2}\f{\hbar^{2}}4
\,,
\ee
where we clearly see that that the coupling in front of the effective potential acquires a new $(2k+1)^{2}$ factor coming from the uncertainty level $k$.  This can be interpreted as an effective intrinsic  angular momentum potential in $(2k+1)\hbar/2$ entirely fuelled  by quantum uncertainty.
We stress that only the pure Gaussian wave-packet case $k=0$ had been studied up to now.

\medskip

The second method towards an effective wave-packet dynamics is to plug the extended Gaussian ansatz directly in the Schr\"odinger field action,
\beq
S[\psi]&=&
\int \left[i\hbar \bpsi\pp_{t}\psi-\f{\hbar^{2}}{2m}\pp_{x}\bpsi\pp_{x}\psi\right]
\nn\\
%&=&
%\int \left[
%Qp\dot{q}-\hbar Q\dot{\gamma}+\hbar \left(k+\f12\right)Q\f{\dot{\alpha}}{2\sigma}+i\f\hbar2 \dot{Q}
%-\f{p^{2}}{2m}-\f{u^{2}}{2m}-\f{\hbar^{2}(2k+1)^{2}}{8m}\f1{w^{2}}
%\right] \\
&=&
\int \left[
Qp\dot{q}-\hbar Q\dot{\gamma}+\f Q2(\beta\dot{\alpha}-\alpha\dot{\beta})+i\f\hbar2 \dot{Q}
-
\left(
\f{p^{2}}{2m}+\f{\beta^{2}}{2m}+\f{\hbar^{2}(2k+1)^{2}}{8m}\f1{\alpha^{2}}
\right)
\right]
\eeq
We recognize the kinetic terms involving time derivative, with total derivative terms, and the Hamiltonian with the kinetic energy contribution in $p^{2}$ and $\beta^{2}$, as well as the conformal potential in $\alpha^{-2}$ for the position spread.

The $Q$ factor might seems unusual. As mentioned earlier, it encodes the normalization of the wave-function, $\int |\psi|^{2}$, i.e. the total number of particles carried by the wave-function.
The equation of motion, resulting from the stationarity of the Shr\"odinger action with respect to global phase variations $\delta\gamma$, clearly imposes that $Q$ is conserved, $\dot{Q}=0$. If we set $Q=1$, as usual when normalizing the wave-function, then we would recover the standard effective action for (free non-relativistic) quantum mechanics.
%\be
%S[\psi]=
%\int \left[
%p\dot{q}+u\dot{w}
%-
%\left(
%\f{p^{2}}{2m}+\f{u^{2}}{2m}+\f{\hbar^{2}(2k+1)^{2}}{8m}\f1{w^{2}}
%\right)
%\right]
%\ee
%with the new factor $(2k+1)^{2}$ in the conformal potential.
%%
%{\bf COMMENT on intrinsic angular momentum coming from quantum dressing of classical particle.}
%
Nonetheless, even without fixing $Q$ to a particular value, one can proceed to the Hamiltonian analysis, introducing conjugate momenta,
\be
P=Qp
\,,\quad
B=Q\beta
\,,\quad
\pi_{Q}=\hbar\gamma+\f12uw
\,,\qquad
\{q,P\}=\{\alpha,B\}=\{Q,\pi_{Q}\}=1\,,
\ee
%\be
%\{q,P\}=\{\alpha,B\}=\{Q,\pi_{Q}\}=1\,,
%\ee
and performing the Legendre transform to set the action in Hamiltonian form:
\be
S[\psi]=
\int \left[
P\dot{q}+B\dot{\alpha}+\pi_{Q}\dot{Q}
-
\left(
\f{P^{2}}{2mQ^{2}}+\f{B^{2}}{2mQ^{2}}+\f{\hbar^{2}(2k+1)^{2}}{8m}\f1{\alpha^{2}}
\right)
\right]
\,,
\ee
where the number of particles $Q$ clearly renormalizes the mass (as expected). Since the Hamiltonian does not contain any $\pi_{Q}$ term, it is obvious that $Q$ is conserved along trajectories.

\medskip

The analysis presented here has shown the extended Gaussian wave-packet ansatz, with Hermite polynomial factors, correspond to excitations of the quantum uncertainty depending to the degree $k$ of the polynomial.
These can be interpreted as the ``classical'' counterpart of the eigenvectors for the  discrete spectrum of the quantum uncertainty operator in quantum field theory.

The extended Gaussian ansatz leads to an effective wave-packet dynamics driven by an effective Hamiltonian controlling the evolution of the mean position and momentum, as well as the position and momentum quadratic uncertainties. We have shown that the effective Hamiltonian includes an effective conformal potential, whose coupling grows with the uncertainty excitation level $k$, thereby extending the previous works by various authors focussing on the pure Gaussian ansatz at $k=0$.

An intriguing remark is that the effective potential has a conformal scaling, and thus corresponds to an angular momentum term, which echoes that the result form the previous section \ref{sec:QFT} that the quantum uncertainty operator, after 2nd quantization, becomes an angular momentum operator when acting on 2-particle and 3-particle states. Although this might signal a deeper logic, this could also be a mere coincidence of the two techniques.

%%%%%%%%
\section*{Conclusion \& Outlook}
%%%%%%%%

The work presented here proposes an incremental improvement for effective Hamiltonian dynamics for quantum mechanics as  developed for instance in \cite{ABCV,Prezhdo2000QuantizedHD,Bojowald:2022lbe}. We clarify the relation between the coupling for the effective conformal potential driving the dynamics of the position spread of the wave-function and the quantum uncertainty characterized by the Heisenberg uncertainty inequality. More precisely, we show how extended Gaussian states, i.e. squeezed states with Hermite polynomial pre-factors, allow to excite the uncertainty and thus change the value of this coupling.

But more importantly, the goal of this short paper is to highlight the quantization of the quantum uncertainty when moving up from quantum mechanics (QM) to quantum field theory (QFT). We underlined that the quadratic moments of the wave-function, defining its spread in position and momentum, are well-defined polynomial observables, and thus should be quantized and raised to quantum operators upon a 2nd quantization from QM to QFT.
Indeed, the wave-function is a classical field and is relegated to a semi-classical object in the QFT context. The quantum uncertainty, defined as a field theory observable in terms of the quadratic moments, becomes an operator in QFT, just as the number of particles.
Using these remarks as starting point, we showed that the  2nd-quantized quantum uncertainty operator has a discrete spectrum given by the Casimir of the $\sl(2,\R)$ Lie algebra for the discrete positive series of unitary representations.
Then squeezed coherent states, with discrete values of the quantum uncertainty, are identified as the QM semi-classical shadows of the discrete spectrum of the uncertainty, as  analogues of Bohr-Sommerfeld trajectories for a quantized Hamiltonian operator.

In non-technical words, since uncertainty reflects the data and information content of the quantum state, the discrete spectrum of the quantum uncertainty would be a further confirmation that information in quantum field theory comes in discrete packets, or in short that information is quantized.
On the theoretical and mathematical side, to solidify the presented results, we need to check the legitimacy of the quadratic moments and uncertainty as well-defined quantum operators acting on the QFT Fock space. On the phenomenological front, we need to understand how to prepare and experimentally realize entangled field wave-functionals such as the eigenstates of the quantum uncertainty identified in the present work.
A third route of possible development is to analyze and identify the QFT fluctuations of the quantum uncertainty, and understand how they affect the effective dynamics of Gaussian wave-packets and squeezed states, most likely generating corrections to the effective potential. It seems that this could be worked out in a similar fashion  as the effect of QFT fluctuations on geodesic motion \cite{Parikh:2020kfh}, using the stochastic approach to quantifying the quantum fluctuations over the QFT vacuum.

We can't help but wonder if this could lead table top experiments in quantum information that put the discrete spectrum of the quantum uncertainty operator in the spotlight and thereby highlight the difference between the QM and QFT regimes. This would confirm (or not) the  necessity of a 2nd quantization, even in the non-relativistic regime.
%

%%%%%%%%%%%%%%%%%
\section*{Acknowledgement}

I am very grateful to the two reviewers of Annals of Physics for helping me clarify the motivation and purpose of the present work.

%%%%%%%%%%%%%%%%%
%\appendix

%%%%%%%%
%\section{}
%%%%%%%%

%%%%%%%%%%
% BIBLIOGRAPHY

%\nocite{*}

\bibliographystyle{bib-style}
\bibliography{QM}

\end{document}